# CloudMonitor: Profiling Power Usage

James William Smith    Ali Khajeh-Hosseini    Jonathan Stuart Ward    Ian Sommerville

School of Computer Science
University of St Andrews
St Andrews, UK
[james.w.smith, ak562, jw497, ian.sommerville]@st-andrews.ac.uk

*Abstract* – **In Cloud Computing platforms the addition of hardware monitoring devices to gather power usage data can be impractical or uneconomical due to the large number of machines to be metered. CloudMonitor, a monitoring tool that can generate power models for software-based power estimation, can provide insights to the energy costs of deployments without additional hardware. Accurate power usage data leads to the possibility of Cloud providers creating a separate tariff for power and therefore incentivizing software developers to create energy-efficient applications.**

***Cloud computing, energy efficient computing, power metering, resource monitoring, cost modelling***

## I. INTRODUCTION

*CloudMonitor*[1], an open-source, automated, scalable, resource and power-usage reporting tool, can provide fine-grain utilization information. The tool includes a power model generation facility to provide software-based energy measurements based on resource usage. This facility provides an accurate estimation of power consumption that can be used to precisely calculate the financial cost of electricity to power a deployment, and opens the possibility of accurate forecasting for cloud providers that may in the future apply a separate tariff for energy consumption.

## II. POWER USE PREDICTION

When calculating the cost of running an IT system, it is common to include the financial cost of electricity under the heading of "*infrastructure*" without fine-grained analysis [1]. Based on historical data, energy costs are rising at 15% per annum. Other infrastructure costs, such as rent or network access, may not rise at the same rate leading to electricity becoming a larger percentage of infrastructure cost.

Power usage data is traditionally obtained by polling Power Distribution Units (PDUs) connected to IT equipment. Dedicated PDU hardware provides accurate data on power consumption but is costly and requires additional maintenance. If the number of machines is increased, a PDU is required for each new machine, making scaling inefficient. Therefore, we seek an alternative means of calculating power consumption through software.

CloudMonitor provides this capability, by initially collecting "training" information from PDUs to ascertain the connections between resource consumption and power used.

This approach is based on work by Bohra & Chaundary in their paper VMeter [2] where they describe a method for predicting power usage of virtual machines by monitoring consumption of hardware resources on the host server - in particular CPU, cache, RAM and hard-drive.

Bohra & Chaundary's work is an improvement over other approaches [3][4] to energy measurement for Cloud environments, as they are not polling dedicated energy monitoring hardware. Their proposed power model is based upon the linear relationship between the sub-components of the system.

However, the weights in the power model of VMeter [2] are workload specific and are calculated manually for each individual application. CloudMonitor automatically analyses resource consumption to create models that are applicable for the current server configuration under any workload.

As machines in a data center are normally procured in batches of the same configuration, training of the model is only required on one machine per batch. The resulting model is able to predict power usage across the remaining servers without the need for additional dedicated metering hardware. If the hardware configuration is the same across multiple machines then the power model is applicable even for different workloads. Our power model is given as follows:

$$Power = \alpha_{Baseline\ Power} + (\beta_1 * CPU) + (\beta_2 * Memory) + (\beta_3 * Hard\ Disk) + (\beta_4 * Network)$$

This model takes into account each hardware sub-component that we measure and generates the weights automatically during the training phase.

### A. Experiment

We developed a video sharing web application[2] as our test system, and developed a client-side tool[3] that mimics users. This application was selected as it represents a typical web application that requires front-end nodes that run web servers, a storage node, and several worker nodes that perform any back-end processing. Our experimental setup for the application used four 2010 Dell PowerEdge R610 servers.

The web server stores uploaded video clips onto the storage server and creates an entry in a queue with the job details of the video that is to be processed. The worker servers perform the video conversion. In order to provide a realistic workload for the application, a tool was created to simulate the browsing behavior of many users accessing the application. The tool performs typical user behavior for a video sharing website - uploading videos, viewing the list of recently videos and watching available clips.

---

[1] https://github.com/jws7/CloudMonitor

[2] https://github.com/alikhajeh1/web_app_experiment

[3] https://github.com/alikhajeh1/web_app_experiment_client

*B. Generating a Power Model*

Performing a linear regression analysis on 24 hours worth of application usage data collected by CloudMonitor results coefficients in Table 1:

TABLE 1: POWER MODEL COEFFICIENTS FOR EXPERIMENT DATA

| Coefficient | Model | Value |
|---|---|---|
| Baseline Power | $\alpha$ | 107.5 |
| CPU | $\beta_1$ | 124.9 |
| Memory | $\beta_2$ | $5.471 \times 10^{-06}$ |
| Hard Disk | $\beta_3$ | $3.661 \times 10^{-02}$ |
| Network | $\beta_4$ | $3.382 \times 10^{-08}$ |

Each of the independent variables in the model had p-values of less than $2 \times 10^{-16}$, suggesting high significance in the equation.

*C. Evaluating the Accuracy of the Power Model*

The intercept of the linear regression correlates strongly with the observed idle power consumption of these particular servers, confirming that it is the value of the *baseline* power. The *dynamic* power consumption above this value is dependent upon the amount of work applied.

To evaluate this power model, we repeated the above experiment; using the same hardware (as required by the premise of a power model relating to a particular hardware configuration) with the workload on each individual machine varied to test the robustness of the model. The generated power model resulted in an average error rate of just 3.91% when applied to the evaluation experiment data.

## III. ENERGY TARIFF

The ability to accurately measure power without costly hardware opens up the possibility for cloud service providers to create a tariff for energy. Explicitly exposing the cost of energy to software developers would encourage them to reduce the energy footprint of their applications.

The video processing web application needs 15.73kWh every day to operate for the workload levels applied. The University of St Andrews is charged a value of $0.14/kWh by their energy provider. Based on historical data, this value is rising at 15% per annum due to rising costs of fuel. At the current rate, including the 15% annual increase, the overall cost for electrical power over a 36-month period would be estimated at $2,767 for this application.

As part of our previous work [5], we have developed tools to support decision makers during the adoption of cloud computing in their organizations. One tool is a cost-modeling application that is available at www.ShopForCloud.com. The data gathered during the experiment can be input into this tool to get a cost estimate of deploying the application on a public cloud, in addition to a breakdown of the cost categories.

Table 2 shows the total cost of the video processing application over a 3-year period, for a hypothetical public cloud that uses a similar pricing scheme as the AWS EU cloud but with an additional charge for energy usage.

TABLE 2: 3-YEAR COST OF VIDEO PROCESSING WEB APPLICATION ON HYPOTHETICAL CLOUD THAT HAS AN ENERGY USAGE TARIFF

| Category | Cost |
|---|---|
| Data transfer | $58,084 |
| Virtual machine hours | $40,568 |
| ***Energy usage*** | ***$2,767*** |
| Storage | $2,325 |
| Storage I/O requests | $2,293 |
| Total | $106,037 |

The energy usage would account for 2.6% of the total costs, a higher value than storage and storage I/O request costs. These values shows that by using a tool such as CloudMonitor, cloud providers can, if they choose to, charge users for the energy consumption of their deployments.

The costs assume current AWS EU pricing for a similar specification to the physical servers that were used in this study. For a more detailed breakdown of the experiment and cost calculations, please see the CloudMonitor website at http://jws7.net/cloudmonitor.

## IV. CONCLUSION

CloudMonitor has been presented as a distributed monitoring infrastructure suitable for gathering information about system deployments. It is able to provide precise information about resource usage and can predict power usage once a power model has been trained on a particular hardware configuration. We have improved on the related work by automatically creating an accurate power model using a live linear regression approach. This automated approach shows a mean accuracy value of 96.09% when evaluating different workloads on the same hardware.

By employing a tool like CloudMonitor, cloud providers can create a new energy tariff for their services and incentivize software developers to create energy efficient applications.